\documentclass[letterpaper,twocolumn,10pt]{article}
\usepackage{usenix}
\usepackage{bbm}
\usepackage{tikz}
\usepackage{amsmath}
\usepackage{booktabs}  %
\usepackage{siunitx} 
\usepackage{amssymb}

\DeclareUnicodeCharacter{2212}{-}

\usepackage{filecontents}
\usepackage{float}  %
\usepackage{dblfloatfix} %
\usepackage[ruled, linesnumbered, boxed]{algorithm2e}  

\begin{document}

\date{}

\title{\Large \bf HYDRA: Unearthing ``Black Swan'' Vulnerabilities in LEO Satellite Networks}

\author{
{\rm Bintao Yuan}$^{1,2}$, {\rm Mingsheng Tang}$^{1,2}$, {\rm Binbin Ge}$^{3}$, {\rm Hongbin Luo1}$^{1,2,3}$, {\rm Zijie yan}$^{4}$\\
$^{1}$School of Cyber Science and Technology, Beihang University\\
$^{2}$Key Laboratory of Space-Air-Ground Integrated Network Security\\
$^{3}$Zhongguancun Laboratory\\
$^{4}$School of Computer Science and Technology, Beihang University
}

\maketitle
\begin{abstract}
As Low Earth Orbit (LEO) become mega-constellations critical infrastructure, attacks targeting them have grown in number and range. The security analysis of LEO constellations faces a fundamental paradigm gap: traditional topology-centric methods fail to capture systemic risks arising from dynamic load imbalances and high-order dependencies, which can transform localized failures into network-wide cascades. To address this, we propose HYDRA, a hypergraph-based dynamic risk analysis framework. Its core is a novel metric, Hyper-Bridge Centrality (HBC), which quantifies node criticality via a load-to-redundancy ratio within dependency structures. A primary challenge to resilience: the most critical vulnerabilities are not in the densely connected satellite core, but in the seemingly marginal ground-space interfaces. These are the system’s ``Black Swan'' nodes—topologically peripheral yet structurally lethal. We validate this through extensive simulations using realistic Starlink TLE data and population-based gravity model. Experiments demonstrate that HBC consistently outperforms traditional metrics, identifying critical failure points that surpass the structural damage potential of even betweenness centrality. This work shifts the security paradigm from connectivity to structural stress, demonstrating that securing the network edge is paramount and necessitates a fundamental redesign of redundancy strategies.

\end{abstract}

\section{Introduction}
The integration of Space-Air-Ground Integrated Networks (SAGIN) has positioned Low Earth Orbit (LEO) satellite networks (LSNs) as a transformative force in global telecommunications \cite{pachlerUpdatedComparisonFour2021,handley2018}. Unlike Geostationary orbit satellite (GEO) systems, LEO constellations such as Starlink, OneWeb, and Kuiper utilize thousands of satellites orbiting at high velocities to provide low-latency and high-bandwidth services \cite{bhattacherjee2019}. These networks operate as complex cyber-physical heterogeneous systems, governed by strict orbital mechanics and resource constraints, which makes their structural analysis fundamentally different from that of terrestrial networks.

As LSNs become critical infrastructure, attacks targeting them have grown in number and range \cite{yue2023}. A primary challenge to resilience is cascading failures\cite{motter2002}, in the resource-constrained environment of LSNs, a localized disruption—caused by targeted cyberattacks, software/hardware failures, space debris, or other factors—can cascade into a network-wide service collapse. To mitigate this, network defense strategies typically rely on identifying critical nodes using topological centrality metrics like degree, betweenness, and pagerank \cite{freeman1978, page1999}. 

\begin{figure}[!htbp]
\centering
\includegraphics[width=0.8\columnwidth, keepaspectratio, trim=2mm 2mm 2mm 2mm, clip]{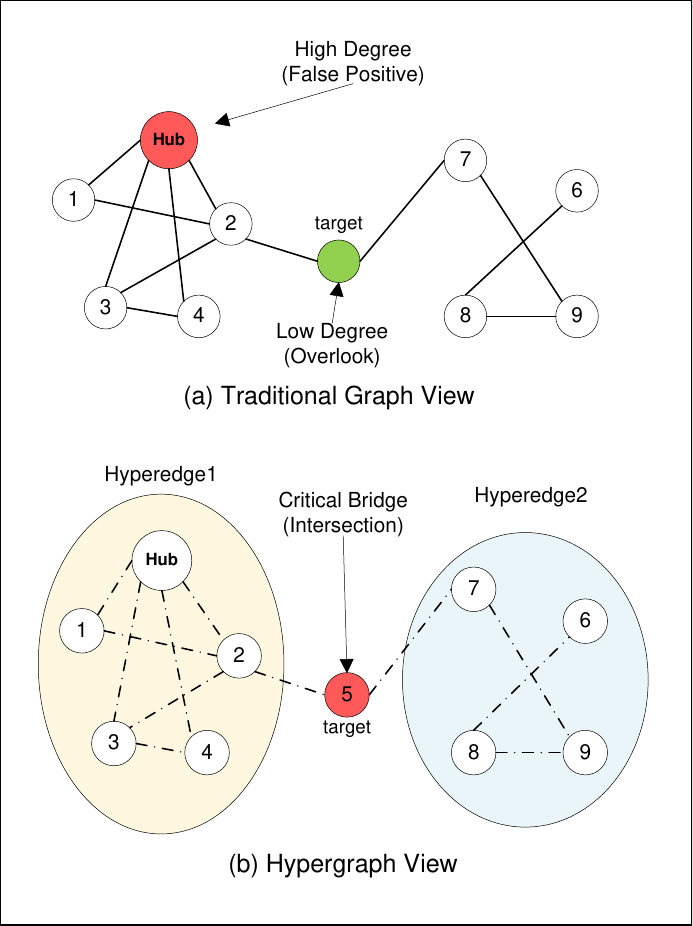}
\caption{\label{motivation} A motivating example of mismatch between topological centrality and structural criticality}
\end{figure}

However, we contend that these traditional metrics are inadequate when applied to the the resource-constrained and dynamic nature of LSNs. 
We present an example in Figure~\ref{motivation}. Consider a ground-space interface node $v_{target}$, in the conventional simple graph view (Figure~\ref{motivation}a), this node possesses a low degree ($k=2$) and is categorized by metrics like degree and betweenness centrality as a peripheral leaf node. In stark contrast, the hypergraph abstraction (Figure~\ref{motivation}b) unmasks its true lethality: $v_{target}$ serves as the sole intersection bridging two functional sets. Vulnerability in LSNs is often driven by structural stress rather than popularity. A node may handle relatively low traffic volume but serve as the sole bridge between two large network partitions. Traditional metrics often fail to capture them, leaving such ``hidden'' critical nodes unprotected.

In this paper, we introduce HYDRA (hypergraph-based dynamic risk analysis), a comprehensive framework designed to detect hidden systemic risks in LSNs. At the core of HYDRA is a novel metric, Hyper-Bridge Centrality (HBC), which quantifies the cascading risk of individual nodes based on load-capacity physics rather than pure topology. HYDRA utilizes a dynamic hypergraph model to capture high-order interactions between satellites, beams, and ground gateways.

Our empirical examination of the Starlink network topology reveals a counter-intuitive finding: HBC exhibits no discernible correlation with traditional metrics, as evidenced by a near-zero Pearson correlation coefficient. This indicates that HYDRA identifies a distinct class of critical nodes that are overlooked by conventional centrality analysis. Specifically, while traditional metrics focus on the dense LSNs core, HBC highlights extreme vulnerabilities at the network edges, particularly in feeder links (gateways). These nodes, which serve as high-stress interfaces between the ground and space segments, act as the systemal bottlenecks.

In summary, this work makes three core contributions:
\begin{itemize}
    \item First, we present HYDRA, a novel and extensible simulation framework that advances the state of the art by synthesizing real orbital mechanics, hypergraph-based dependency modeling, and load-aware cascade simulation. It is the first to model the heterogeneous, time-varying dynamics of LEO mega-constellations, enabling high-fidelity security analysis.
    \item  Second, we establish HBC as a core analytical innovation. This metric redefines node criticality by measuring structural stress instead of topological centrality, offering a principled method to uncover systemic vulnerabilities that conventional metrics miss.
    \item Third, we demonstrate, via an empirical study of Starlink, the critical existence of ``Black Swan'' vulnerabilities. Our results validate that an HBC-informed strategy substantially outperforms topology-based baselines, conclusively identifying structural stress—not connectivity—as the paramount factor for risk assessment in dynamic, resource-constrained networks.
\end{itemize}

\section{Background}

This section establishes the theoretical foundations required to analyze systemic vulnerabilities in LSNs. We detail the orbital mechanics governing topology dynamics, the method of routing and switching, and the mathematical necessity of hypergraph modeling. Finally, we characterize the geographic heterogeneity of internet traffic.
\subsection{LSNs Orbital Dynamics}
Temporal variability is the defining characteristic of LSNs. Unlike GEO satellites or terrestrial base stations, LEO satellites operate at altitudes of 500–1,200 km with orbital velocities approximating 7.6 km/s. This results in highly dynamic connectivity, where the visibility window for a ground user is typically limited to 3–5 minutes \cite{handley2018}.

To achieve high-fidelity simulation of this dynamic topology, we utilize the Simplified General Perturbations-4 (SGP4) propagator  \cite{valladoSGP4OrbitDetermination2008}, the industry standard for processing North American Aerospace Defense Command (NORAD) Two-Line Element (TLE) sets \cite{CelestrakNORADElements}. SGP4 accounts for non-spherical Earth gravitational effects ($J_2$ perturbation) and atmospheric drag, enabling precise calculation of the satellite state vectors $\vec{r}(t)$ and $\vec{v}(t)$. In our HYDRA framework, these calculations are performed at minute-level granularity to strictly enforce line-of-sight (LoS) constraints for ISLs formation.
\subsection{ISLs Routing and Handover Management}
Due to the relative motion of satellites, the connectivity status of ISLs is highly dynamic. Consequently, LSNs are formally modeled as discrete time-varying graph, where the continuous timeline is discretized into a sequence of static snapshots.

Regarding routing, and given the low-latency objective of LSNs architectures, traffic forwarding predominantly adheres to the shortest path first principle. However, this snapshot-based routing faces the challenge of capacity rigidity: satellite link bandwidth is strictly limited by power budgets and Shannon limits, lacking the elastic buffering capabilities of terrestrial fiber networks. LSNs will be affected when topological mutations or traffic surges occur.

These orbital dynamics and routing constraints collectively create a fragile foundation where localized congestion can rapidly propagate, as explored in our cascade model (Section \ref{sec:cascade_model}).
\begin{figure}[tbp]
\centering
\includegraphics[width=0.8\columnwidth, keepaspectratio, trim=1mm 1mm 1mm 1mm, clip]{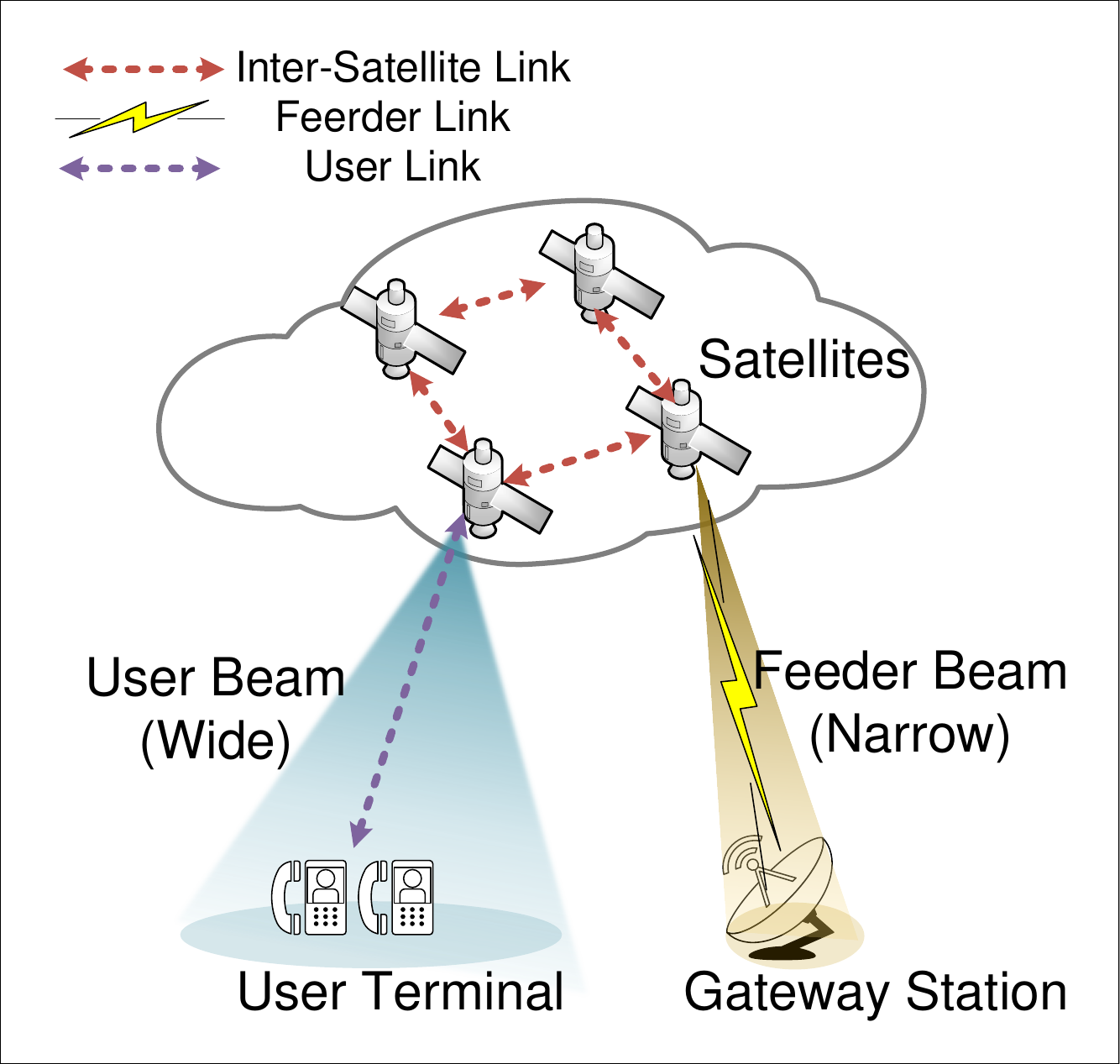}
\caption{\label{LSN} Basic transmission schematic of LSNs}
\end{figure}

\subsection{Hypergraph Theory and Beam Coverage}
Traditional complex network analysis models the system as a simple graph $G=(V, E)$, where each edge $e \in E$ connects exactly two vertices. We argue that this binary abstraction creates a semantic gap when modeling LSNs cascade failure. In LSNs architectures, a single satellite projects multiple spot beams, each serving hundreds of users or gateways simultaneously. This creates a one-to-many dependency: if a specific beam is disabled, all users within its footprint are affected (Figure~\ref{LSN}). Simple graph models fail to capture this high-order interaction, often underestimating the impact of edge-layer failures. To address this, we employ hypergraph theory \cite{bretto2013}. We define the failure network as a hypergraph $H=(V, \mathcal{E})$, where a hyperedge $e \in \mathcal{E}$ can connect an arbitrary number of vertices ($|e| \ge 2$). This formulation allows HYDRA to rigorously quantify the bulk disconnection risks that propagate from the physical layer (satellite) to the logical layer (beam) and finally to the access layer (user).
\subsection{Internet Traffic Distribution}
Real-world network traffic is not uniformly distributed across nodes; rather, it exhibits strong geographic heterogeneity. Ignoring this distribution leads to biased centrality assessments.

Internet traffic generation follows a gravity model, constrained by the topology of autonomous systems (AS) \cite{roughan2005}. According to CAIDA datasets, global traffic demands are heavily skewed towards densely populated regions and submarine cable landing stations, while oceanic and polar regions carry minimal source/sink traffic. In this study, we synthesize traffic matrices based on gridded population of the world  data and AS-level interconnectivity \cite{NASA-GPW}. This ensures that our HBC metric reflects actual load stress rather than theoretical topological popularity, explaining our finding that low-degree nodes bridging high-demand regions are often more critical than high-degree nodes.
\section{Overview}
This section defines the adversarial model considered in our study, presents the architectural design of the  HYDRA framework, and formalizes the concept of ``Black Swan'' vulnerabilities in the context of LSNs.

\subsection{Threat Model}
Adversary Goals: The primary objective is to maximize service unavailability  and trigger widespread cascading failures. The adversary seeks to achieve this by targeting a minimal set of nodes.

Adversary Capabilities: We assume the adversary possesses global visibility of the network topology, derived from publicly available TLE data and orbital mechanics. We define an attack as an increase in node load; the attack intensity is determined by the parameter $\alpha$ in the equation (\ref{eq:tilde_Ci}), and a node will be removed when its load exceeds a certain threshold.

This attack definition corresponds to realistic attack vectors: including firmware exploitation on orbital nodes that facilitates blackhole routing or logical beam disabling, as well as terrestrial assaults via high-power uplink jamming or volumetric DDoS attacks on critical ground-space gateways. The adversary is budget-constrained, capable of compromising at most $k$ nodes, where $k \ll N$ ($N$ is the total network size).

Defense Assumptions: The network operator employs shortest path first routing based on real-time topology snapshots. We assume the links and nodes have hard capacity thresholds (e.g., throughput limits). No massive over-provisioning is assumed, reflecting the economic constraints of commercial constellations.

\subsection{HYDRA Framework Architecture}
\begin{figure*}[!t]
\centering
\includegraphics[width=\textwidth,height=0.50\textheight,keepaspectratio,trim=5 25 5 5,clip]{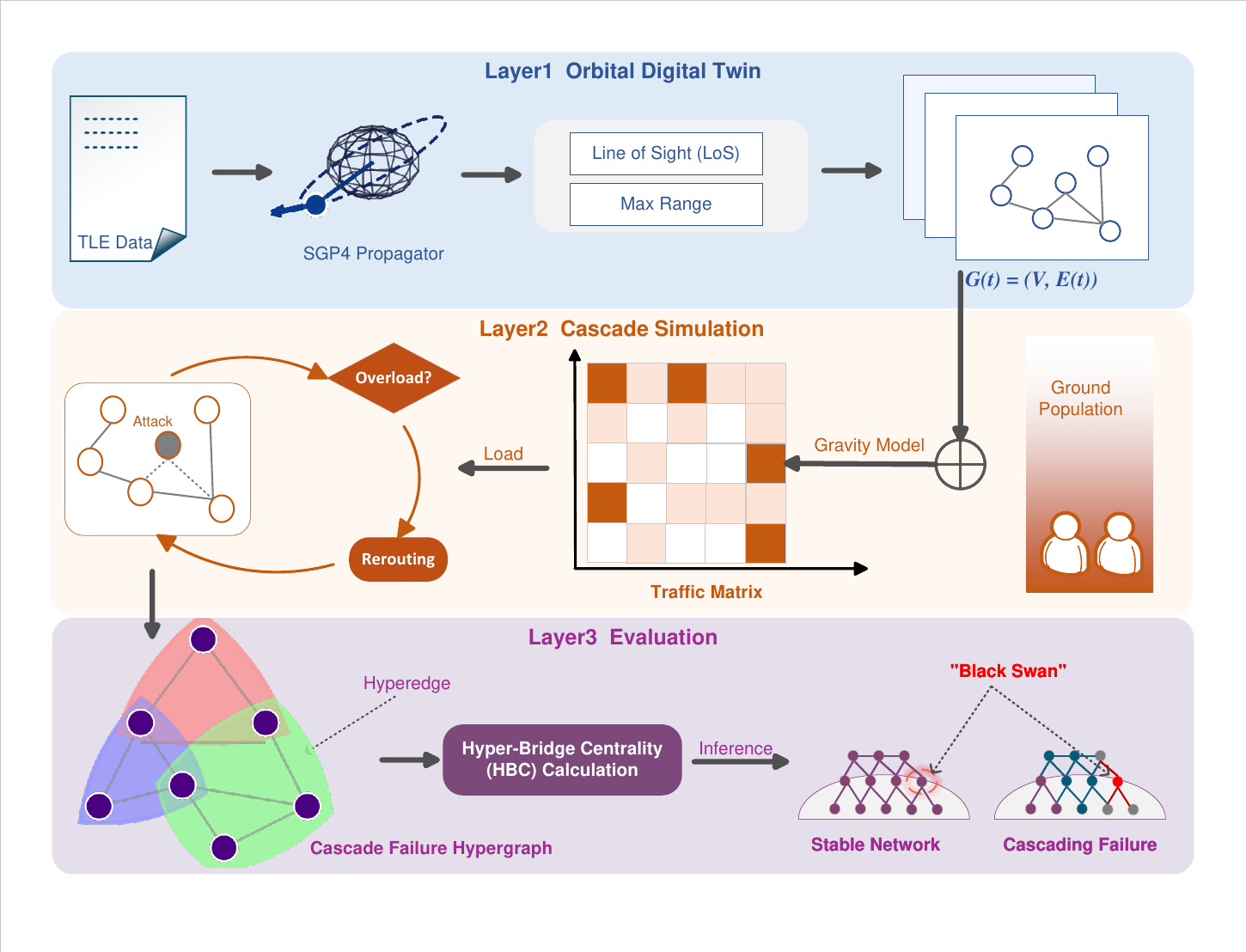}
\caption{\label{r2} HYDRA framework architecture}
\end{figure*}
As illustrated in Figure~\ref{r2}, the framework operates as a three-layer pipeline:

Layer 1: Orbital Digital Twin. This layer establishes the physical foundation of the simulation. It utilizes an SGP4 Propagator to process TLE data, strictly applying Physical Constraints such as Los and max range. This process generates a high-fidelity time-varying graph, outputting a sequence of discrete topology snapshots that represent the dynamic connectivity of the LSNs over time.

Layer 2: Cascade Simulation. This layer models the propagation of failures. It integrates the topology with a global traffic matrix, which is synthesized based on population density and inter-regional flows (e.g., North America (NA), Europe (EU), Asia (AS) ). The simulation is triggered by initialization capacity and injection attack scenarios. It executes a recursive failure loop: node removal  based on attack targets, rerouting  of traffic flows, overload assessment where ``Load > Capacity''. This cycle continues until the network stabilizes, resulting in a distinct failure set.

Layer 3: Evaluation. The final layer performs the high-order theoretical analysis. It transforms the physical network into a hypergraph abstract  to capture complex dependencies. Specifically, it constructs a hyperedge to encapsulate the relationships between a satellite, its beams, the gateway, and the users. Based on this structure, the framework performs the HBC calculation. This metric drives the final risk analysis and allows for precise bottlenecks identify within the constellation.
\subsection{The Black Swan Node}
Based on the recent theoretical framework by Zhu et al.\cite{zhu2025}, we formally define the specific vulnerability HYDRA aims to detect.

Definition: A node $v$ in an LSNs is defined as a black swan node if it satisfies two conditions simultaneously:

\textbf{(i) Topological Insignificance.} Its degree centrality $d(v)$ is below a trivial threshold $\delta$ (i.e., it appears unimportant in the static graph).
$$ d(v) < \delta$$
   
\textbf{(ii) Structural Lethality.} Its HBC exceeds a critical stress threshold $\tau$ (i.e., it is essential for handling overflow traffic).
$$ HBC(v) > \tau$$
   
Problem Statement: Traditional metrics effectively identify nodes where $d(v) \to \text{high}$. The goal of HYDRA is to unearth the set of nodes $V_{BS} = \{v \mid d(v) < \delta \land HBC(v) > \tau\}$, which act as ``hidden bridges'' capable of triggering disproportionate cascading failures upon removal.
\section{Methodology}
Our methodology integrates three core components: (i) a high-fidelity, time-varying LSNs model capturing orbital dynamics and hierarchical architecture; (ii) a physics-aware cascade simulation engine modeling load redistribution; and (iii) a hypergraph-based analysis framework for extracting systemic risk metrics. We detail each component below.             
\subsection{LSNs System Model}
Unlike conventional terrestrial network, LSNs is characterized by high mobility and intermittent connectivity. We model the system as a dynamic, heterogeneous graph driven by orbital mechanics.
\subsubsection{Multi-layer Node Abstraction}
We model the network as a time-varying graph,
\begin{equation}
G(t) = (\mathcal{V}, \mathcal{E}(t))
\end{equation}
where $\mathcal{V}$ is the node set and $\mathcal{E}(t)$ is the edge set at time $t$. The node set $\mathcal{V}$ is defined as:
\begin{equation}
\mathcal{V} = \mathcal{S} \cup \mathcal{B} \cup \mathcal{G} \cup \mathcal{U}
\end{equation}
Here, $\mathcal{S}$ represents satellites, $\mathcal{G}$ represents ground gateways, and $\mathcal{U}$ represents user terminals. A key modeling insight is the explicit representation of satellite beams as logical access nodes (set $\mathcal{B}$). In LSNs, beams act as the true access and aggregation entities that concentrate user demand, and users attach logically to beams rather than directly to the satellite’s core forwarding fabric (figure ~\ref{LSN}). In our model, users attach to beams ($u \to b$), and beams attach to satellites ($b \to s$). This distinction is vital because congestion often originates at the beam input/output level rather than at the satellite core. Neglecting the beam layer can therefore systematically reduce the accuracy of risk assessments.

\subsubsection{Time-Varying Connectivity}
The edge set $\mathcal{E}(t)$ is governed by orbital mechanics and physical visibility. For every satellite $s \in \mathcal{S}$, we propagate its position $\mathbf{r}_s(t)$ using the SGP4 model:
\begin{equation}
r_s(t) = \text{SGP4}(s, t)
\end{equation}
SGP4 is an efficient satellite orbit propagator capable of achieving moderately accurate and rapid orbit predictions based on TLE data, which is consistent with operational satellite tracking practice~\cite{valladoSGP4OrbitDetermination2008}.

Connectivity between any two nodes $i, j$ is defined by a binary visibility function $V_{ij}(t)$, which accounts for the maximum communication range $D_{ij}^{\max}$ and geometric constraints (such as minimum elevation angles and earth occlusion):
\begin{equation}
V_{ij}(t) = \begin{cases}  1, & \text{if } \|\mathbf{r}_i(t) - \mathbf{r}_j(t)\| \le D_{ij}^{\max} \land \text{constraints met} \\ 0, & \text{otherwise} \end{cases}
\end{equation}
Here, $D_{ij}^{\max}$ is dependents on link-type and can incorporate antenna directionality and link budget constraints. Beam coverage is handled similarly: if a user $u$ lies within beam $b$'s footprint at time $t$, then $V_{bu}(t) = 1$.

To optimize simulation efficiency for mega-constellations, we employ an ``Active Subgraph Selection'' strategy: only satellites visible to active ground users and their multi-hop neighbors are dynamically loaded into the graph $G(t)$ at each snapshot, ensuring computational feasibility without sacrificing local topological accuracy.

\subsubsection{Capacity of Links and Nodes}
We distinguish between physical link capacity and node processing capacity. This separation is important because cascades in operational networks are often triggered by node-level overload , even when individual links remain nominal.

\textbf{(i)Physical Link Capacity.} For each active link, we compute Signal Noise Ratio(SNR) driven by distance-dependent path loss:
$$
\operatorname{SNR}_{ij}(t)=\frac{P_t G_i G_j}{N_0 B_{ij} L_{\text{path}}(\|\mathbf{r}_i-\mathbf{r}_j\|)}
$$
Where $P_t$ represents the transmitted power, $G$ represents the antenna gain, $N_0$ represents the noise power spectral density, and $B_{ij}$ represents the channel bandwidth.

Based on the Shannon-Hartley theorem, the physical link capacity (maximum throughput) $C_{ij}^{\text{phy}}(t)$ is defined as~\cite{shannonMathematicalTheoryCommunication}:
$$ 
C_{ij}^{\text{phy}}(t)=B_{ij}\log_2\left(1+\text{SNR}_{ij}(t)\right)
$$

In our simulation implementation, to balance fidelity with computational efficiency for large-scale constellations, we approximate these continuous values. We utilize parameterized capacity thresholds derived from standard link budget analyses for optical ISLs and RF user beams, treating them as bounded constraints rather than computing real-time fading for every packet.

\textbf{(ii) Node Effective Capacity.} The processing capacity of a node is not equivalent to its link capacity. Beyond radio constraints, the capacity of satellites and beams is fundamentally constrained by onboard hardware resources, such as switching fabric and buffer size. We define the effective service capacity $C_i$ for a node $i$:
\begin{equation}
C_s^{\text{tot}} = \min \left( C_s^{\text{hw}}, \sum_{j: V_{sj}(t)=1} C_{sj}^{\text{phy}}(t) \right)
\end{equation}

This formulation ensures that a node cannot process more traffic than its hardware allows, regardless of how many links it possesses. Thus, the total capacity of the node is:
\begin{equation}
C_i =
\begin{cases}
    C_s, & i \in \mathcal{S} \\
    C_b, & i \in \mathcal{B} \\
    C_g, & i \in \mathcal{G}
\end{cases}
\end{equation}
$C_b$ represents the effective capacity of the beam node, $C_s$ represents the total satellite transponder processing capacity, $C_g$ represents the ground station capacity.
\subsubsection{Traffic Model}				
In LSNs, topology changes do not immediately lead to collapse, whereas significant traffic fluctuations pose a more immediate risk to network stability. Therefore, In our work, the communication carrier is defined as the business traffic, rather than the state of the nodes.

We generate user–gateway traffic with a lightweight demand model consistent with the simulator. Each user $u \in U$ generates a flow $f_u = (u, g(u), d_u)$, where the demand $d_u$ is proportional to the user’s population weight and a temporal factor (local time), and is globally scaled to match the system target load:

\begin{equation}
d_u \propto \text{PopWeight}_u \cdot \phi(\text{lon}_u, t)
\end{equation}
$\phi(\text{lon}_u, t)$  represents the time factor, indicating the user's activity or demand level in local time.
A global scaling factor is then applied so that the total offered load is aligned with available beam capacity and target load.

To map this demand to physical nodes, we follow the same routing logic as the simulator. Each flow is routed by a load‑aware Dijkstra algorithm (edge weights depend on residual capacity), resulting in a single selected path at time $t$. The instantaneous load on node $i$ is the aggregate demand of all flows traversing it:

\begin{equation}
L_i(t) = \sum_{u} \sum_{p \ni i} x_{u,p}(t) \cdot d_u
\end{equation}
where $x_{u,p}(t)=1$ indicates the path selected for flow $f_u$.

\subsection{Flow-Capacity Cascade Dynamics}
\label{sec:cascade_model}
Standard epidemic models (e.g., Susceptible-Infected-Recovered (SIR) model) or abstract state maps (e.g., Cellular Markov Logic (CML) framework) fail to capture the physics of network congestion. In LSNs, failures propagate because the rerouting of traffic from a failed node overloads adjacent nodes. We model this as a discrete-time process driven by deterministic overload and asynchronous rerouting.
\subsubsection{Overload and Failure Condition}
We formulate a cascade model where failures emerge when the rerouted traffic exceeds node service capacity. A node $i$ is considered failed if its instantaneous load exceeds its effective capacity, so we define the overload condition: $L_i(t) > {C}_i$.

To model realistic cyber-physical attacks, we introduce a capacity degradation factor $\alpha$, which captures the effect of partial node compromise (e.g., due to jamming or software exploitation). An attack does not necessarily remove a node but may degrade its capacity (e.g., via resource exhaustion or DoS): 
\begin{equation}
\tilde{C}_i = (1 - \alpha q_i) C_i
\label{eq:tilde_Ci}
\end{equation}
where $q_i$ is the attack/compromise level, determined by the characteristics of the attack method and $\alpha \in [0,1]$ is the severity coefficient, determined by the scale and intensity of the attack.

So the overload condition is updated as follows:
\begin{equation}
L_i(t) > \tilde{C}_i
\label{eq:L_i(t)}
\end{equation}
\subsubsection{The Cascade Simulation Algorithm}
Let $F_k\subset \mathcal{V}$ be the set of failed nodes after iteration $k$. The cascade proceeds as follows:

\textbf{Step 1: Initialization.} $k=0$: an initial set of nodes $F_0$ is disabled or degraded based on the attack scenario.
  
\textbf{Step 2: Topology Update.} Remove $F_k$ from the current graph to obtain the surviving subgraph $G_k = G(t) \setminus F_k$.

\textbf{Step 3: Load-Aware Rerouting.} For all affected flows $(u, g)$, new paths are computed on $G_k$. We employ the load-aware shortest path algorithm:
 \begin{equation}
 p^*_{u,g} = \arg\min_{p} \sum_{\ell \in p} \left( \frac{1}{{C}_\ell - L_\ell^{(k-1)}} + \delta \cdot \text{delay}_\ell \right)
 \end{equation}
$p^*_{u,g}$ represents the shortest path from uere $u$ to gateway $g$, $C_l$ denotes the aggregate capacity of link $l$, representing the maximum traffic flow that the link can accommodate. $L_l^{(k-1)}$ corresponds to the existing traffic load on link $l$ accumulated during the preceding iteration $(k-1)$. So, $C_l - L_l^{(k-1)}$ quantifies the residual capacity of link $l$. $\frac{1}{C_l - L_l^{(k-1)}}$ is defined as the reciprocal of the residual capacity, the higher of the capacity, the lower of this formula. $\text{delay}_l$ denotes the transmission latency associated with link $l$, quantifying the time duration required for data traversal across the specific link, $\delta$ serves as a weighting coefficient designed to modulate the influence of link latency on the  path selection. 

Generally, this formulation aims to identify a routing path that maximizes link capacity and minimizes transmission latency,  ensuring a traffic distribution mechanism that enhances the overall efficiency and stability of the network.

\textbf{Step 4: Load Update and Overload Decision.} Based on the new routes, the load $L_i^{(k)}$ of node $i$ is defined as the aggregate summation of all traffic demands traversing the node:
\begin{equation}
 L_i^{(k)} = \sum_{\text{all flows } f} \mathbbm{1}(i \in \text{path}_f) \cdot d_f
\end{equation}
Any node exceeding its capacity is added to the failure set:
\begin{equation}
F_{k+1} = F_k \cup \{ i \mid L_i^{(k)} > \tilde{C}_i \}
\end{equation}

\textbf{Step 5: Termination.}  The process repeats until $F_{k+1} = F_k$ (system stabilizes) or the network becomes disconnected for the required services.

\subsection{Hypergraph Analysis}
Traditional graph metrics analyze connection relationships and often underestimate systemic risk in complex systems. They fail to capture ``joint failure events'', where a specific group of nodes fails simultaneously due to shared dependencies. To address this, we construct a hypergraph representation of cascade outcomes.
\subsubsection{Constructing the Failure Hypergraph}
We define a cascading-failure hypergraph $\mathcal{H} = (\mathcal{V}, \mathcal{E}_H)$. Each hyperedge $e_m \in \mathcal{E}_H$ represents the `set of all nodes' that failed locally during a specific simulation run $m$:

\begin{equation}
\mathcal{E}_H = \{e_1, e_2, \dots, e_M\}, \quad e_m = F_{\text{final}}^{(m)}
\end{equation}
By aggregating results from Monte Carlo simulations across various attack vectors and traffic patterns, $\mathcal{H}$ captures higher-order correlations between nodes.

\subsubsection{Hypergraph-Based Risk Metrics}
We introduce HBC to identify ``Black Swan'' nodes: components that appear robust (low degree) but trigger catastrophic cascades.

\textbf{(i) Cascading Failure Risk (CFR)}

The CFR of a node $v$ quantifies the expected system-wide damage conditioned on $v$ being part of the initial failure set:

\begin{equation}
\text{CFR}(v) = \mathbb{E} \left[ \frac{|F_{\text{final}}|}{|\mathcal{V}|} \; \middle| \; v \in F_0 \right]
\end{equation}

This measures the effect of a node failure rather than its topological importance.

\textbf{(ii) Hyper-Bridge Centrality (HBC)}

To distinguish hidden risks from obvious bottlenecks, we normalize CFR by the node's physical degree. A high HBC indicates a node that is structurally inconspicuous (low physical degree) but systemically critical:

\begin{equation}
\text{HBC}(v) = \frac{\text{CFR}(v)}{\log(1 + \text{Degree}_{\text{phy}}(v))}
\end{equation}
In this formula, the logarithmic function is employed to adjust the influence of the physical degree, aiming to prevent the results from becoming disproportionate when the degree is excessively large.

Nodes with high HBC values represent the most dangerous vulnerabilities in the network—those likely to be overlooked by standard centrality-based hardening strategies.

\section{Implementation}
\begin{table*}[t]
  \centering
  \caption{Simulation Parameters}
  \label{tab:sim_params}
  \resizebox{\linewidth}{!}{  
  \begin{tabular}{l l l l}
    \toprule
    \textbf{Category}           & \textbf{Parameter}           & \textbf{Symbol}        & \textbf{Value / Configuration}                                \\
    \midrule
    \textbf{Orbital Dynamics}   & Data Source             & ---                    & Starlink TLE (CelesTrak)                                     \\
                                & Propagation Model       & $\mathbf{r}_s(t)$      & SGP4                                                 \\
                                & Network Subset          & $N_{\text{sat}}$       & Active Subgraph from $\approx 9300$ Satellites               \\
                                & Simulation Horizon      & $T_{\text{total}}$     & 90 minutes/ 4320 minutes                                                   \\
                                & Sampling Interval       & $\Delta t$             & 22 minute / 15 minutes                                                    \\
    \textbf{Connectivity}       & Max ISL Range           & $D_{\max}$             & 5,000 km (Configurable)                                       \\
                                & ISL neighbors (k-NN)          & $k_{\max}$             & 4 (Grid-like Topology)                                        \\
                                & Min Elevation Angle     & $\theta_{\min}$        & $25^\circ$                                                   \\
                                
    \textbf{Node Architecture}  & Active Beams            & $N_{\text{beams}}$     & 12 per Satellite                                              \\
                                & User Selection          & $K_{\text{user}}$      & Top-3 Visible Satellites                                      \\
                                & Ground Nodes            & $N_{\text{user}}$      & 800 Users, 8 Citys and 10 Gateways                                         \\
    \textbf{Traffic \& Capacity}& Demand Model            & $T_{ij}$               & Population-based Gravity Model (weighted by AS14593 data)    \\                      
                                & ISL Capacity            & $C_{\text{isl}}$       & 40,000 Mbps (Fixed)                                           \\ 
                                & Node Capacity           & $C_i$                  & Hardware Processing Limit ($C^{\text{hw}}$)                                    \\ 
    \bottomrule
  \end{tabular}
  }
\end{table*}
We developed  a high-fidelity, data-driven simulation environment to realize this theoretical framework. Unlike abstract random graph models, HYDRA reconstructs the network topology at discrete sampling intervals using real-world orbital ephemeris data. This approach captures the temporal variability of LSNs connectivity, meets the computational requirements for extensive cascading failure analysis.                
\subsection{Orbital Dynamics and Topology Generation}
We instantiate the simulation using Starlink TLE sets acquired from CelesTrak~\cite{StarlinkTle}. To manage computational complexity while preserving topological characteristics, we model a dynamic active window from Starlink TLE ($\approx 9300$ satellites). Also, our open-source simulation framework allows researchers to scale up computational quantity as needed.
  
\textbf{(i) Active Node Selection.} We construct a snapshot $G(t)$ containing only satellites visible to ground users and their $k$-hop neighbors (where $k=2$ by default). This expand mode ensures that all relevant routing paths are captured while ignoring disconnected satellites on the other side of the Earth. The detailed implementation of the algorithm can be found in Appendix \ref{algorithm_detail}.

\textbf{(ii) Temporal Resolution.} The system state is sampled at a high-frequency interval of $\Delta t = k$ minute to accurately capture fast-moving LSNs topology changes (e.g., handovers).

\subsection{Link and Node Configuration}
The network topology $\mathcal{E}(t)$ is constructed dynamically at each time step based on physical visibility and hardware constraints.

\textbf{(i) Inter-Satellite Links (ISLs).} A link is established if distance $d_{ij} \le 5000$ km and LOS is not occluded by the Earth. We enforce a neighbor constraint of 4 to mimic the grid-like laser mesh of Starlink Gen2.

\textbf{(ii) Hierarchical Access Layer.} Consistent with our multi-layer node abstraction, we explicitly model the $u \to b \to s$ hierarchy:

Beam Instantiation. Each satellite is modeled with  multiple logical beam nodes representing different azimuth sectors ($b \rightarrow s$). This abstraction captures access-layer congestion and spatial reuse without modeling detailed fast beam-hopping slots.

User Association. Users ($u$) attach to the beam nodes of the  Top-K satellites (default $K=3$) that maximize elevation angle; a minimum elevation of $25^\circ$ is enforced to define visibility. 

\subsection{Real-World Traffic Model}
To simulate a realistic network load $L_i(t)$, we deploy a scaled representation of 800 user nodes (each modeling a user group) and their corresponding gateways across eight major metropolitan areas. Traffic demand is then generated based on demographic data, calibrated against real-world service adoption patterns.

User density is proportional to the population of the specific metropolitan zone according to standard gravity models, and commercial LSNs exhibit significant geographic heterogeneity in service availability and user adoption. Meanwhile, we leverage empirical traffic data from Cloudflare Radar for AS14593 (Starlink) over a 52-week period \cite{StarlinkTrafficData}. This dataset reveals that the majority of traffic originates from North America, Europe, and Oceania, despite their varying population densities. We implement this by assigning a regional adoption coefficient $\beta_r \in [0, 1]$ to each metropolitan area in our simulation.

Consistent with the dataset, our code configures $\beta \approx 1.0$ for North American cities (e.g., New York, Los Angeles), $\beta \approx 0.8$ for European/Oceanian cities (e.g., London, Paris, Sydney), and significantly lower values ($\beta \le 0.3$) for other regions (e.g., Tokyo, Sao Paulo). This calibration ensures that simulated congestion aligns with the physical reality of high-load corridors.

Furthermore, traffic demand is not static but exhibits dynamic spatiotemporal fluctuations driven by human activity cycles. We implement a continuous diurnal activity function $\mathcal{T}(t, \lambda)$ that captures a Gaussian‑shaped load peak centered at 21:00 local time, with wrap‑around at the day boundary. This temporal factor induces “follow‑the‑sun” congestion shifts as peak hours migrate across longitudes, reflecting sun‑synchronous load dynamics in LEO systems.

To stress the backbone under long‑haul conditions, our traffic generator randomly assigns destination gateways (when multiple gateways are available), rather than always selecting the nearest egress. This forces flows to traverse multi‑hop ISLs and exposes structural vulnerabilities of the orbital core rather than only edge‑access congestion.

A detailed summary of the simulation parameters, including orbital settings and physical constraints, is provided in table~\ref{tab:sim_params}.

\section{Evaluation}
We evaluate HYDAR framework on the following research questions:
\begin{description}
  
\item[RQ1:] Can HBC identify ``Black Swan'' Vulnerabilities that traditional metrics overlook in LSNs?
\item[RQ2:] What are the characteristics of HBC nodes?
\item[RQ3:] Are the HBC nodes fixed or time-varying?
\item[RQ4:] Are the HBC nodes effective in cascading attacks?
\end{description}

In our subsequent analysis, we will primarily examine the $\mathcal{S} \cup \mathcal{B} \cup \mathcal{G}$ nodes, while the  $\mathcal{U}$ nodes represent the needs of users and is not included in the LSNs cascade risk assessment.

\subsection{Unearthing Hidden Vulnerabilities (RQ1)}
We first investigate whether the proposed HBC metric can identify critical nodes that are overlooked by classical topology-based metrics. After 5 random sampling calculations within 90 minutes, we compute the average value to  compare the HBC with classical topology-based metrics for all active nodes. 
\begin{figure}[t]
\centering
\includegraphics[width=\columnwidth,keepaspectratio,trim=2 2 2 2,clip]{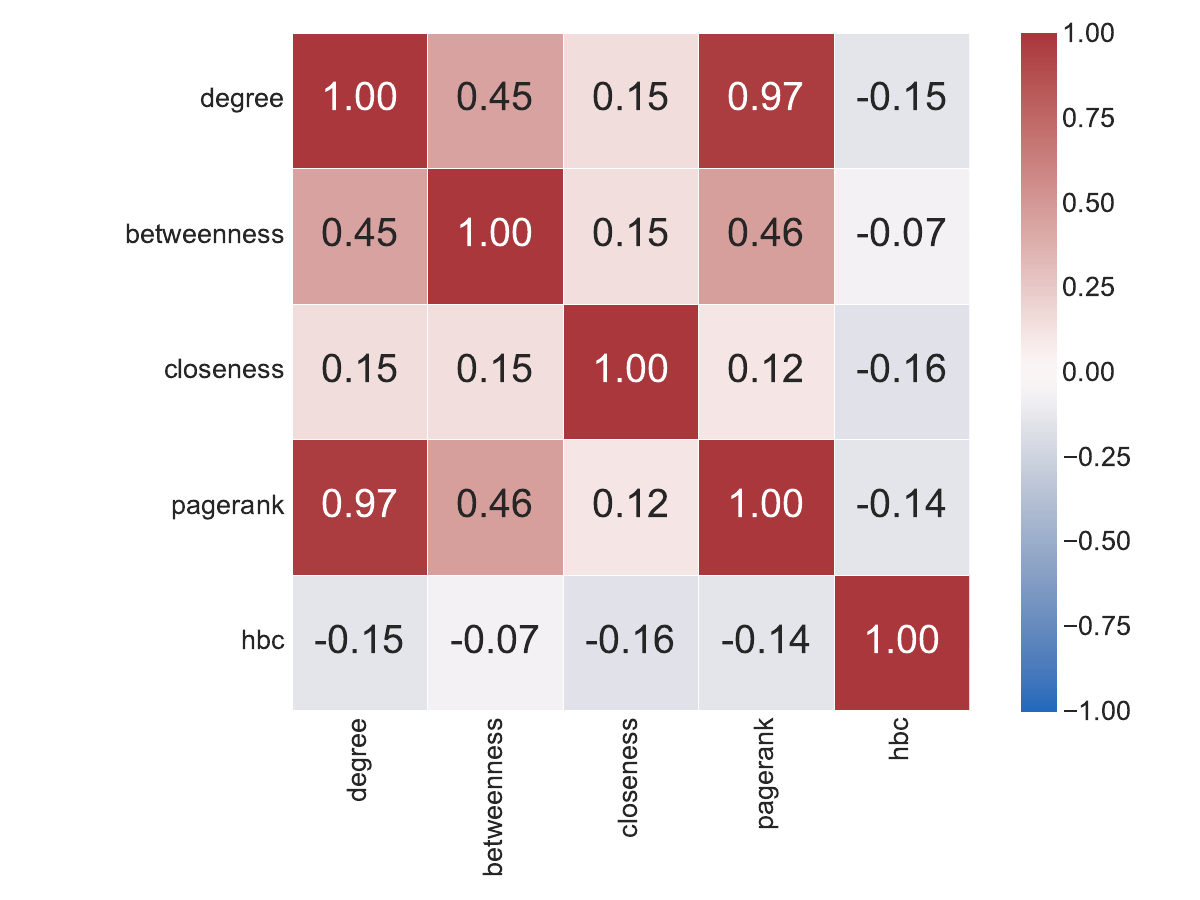}
\caption{\label{rq1} Pearson correlation matrix of metrics}
\end{figure}
Figure~\ref{rq1} presents the Pearson correlation matrix among metrics. A key observation is the strong negative correlation between HBC and degree centrality ($r \approx -0.15$), indicating that nodes critical for cascading risk are distinct from topologically central hubs.

\begin{figure}[t]
\centering
\includegraphics[width=\columnwidth,keepaspectratio,trim=2 2 2 2,clip]{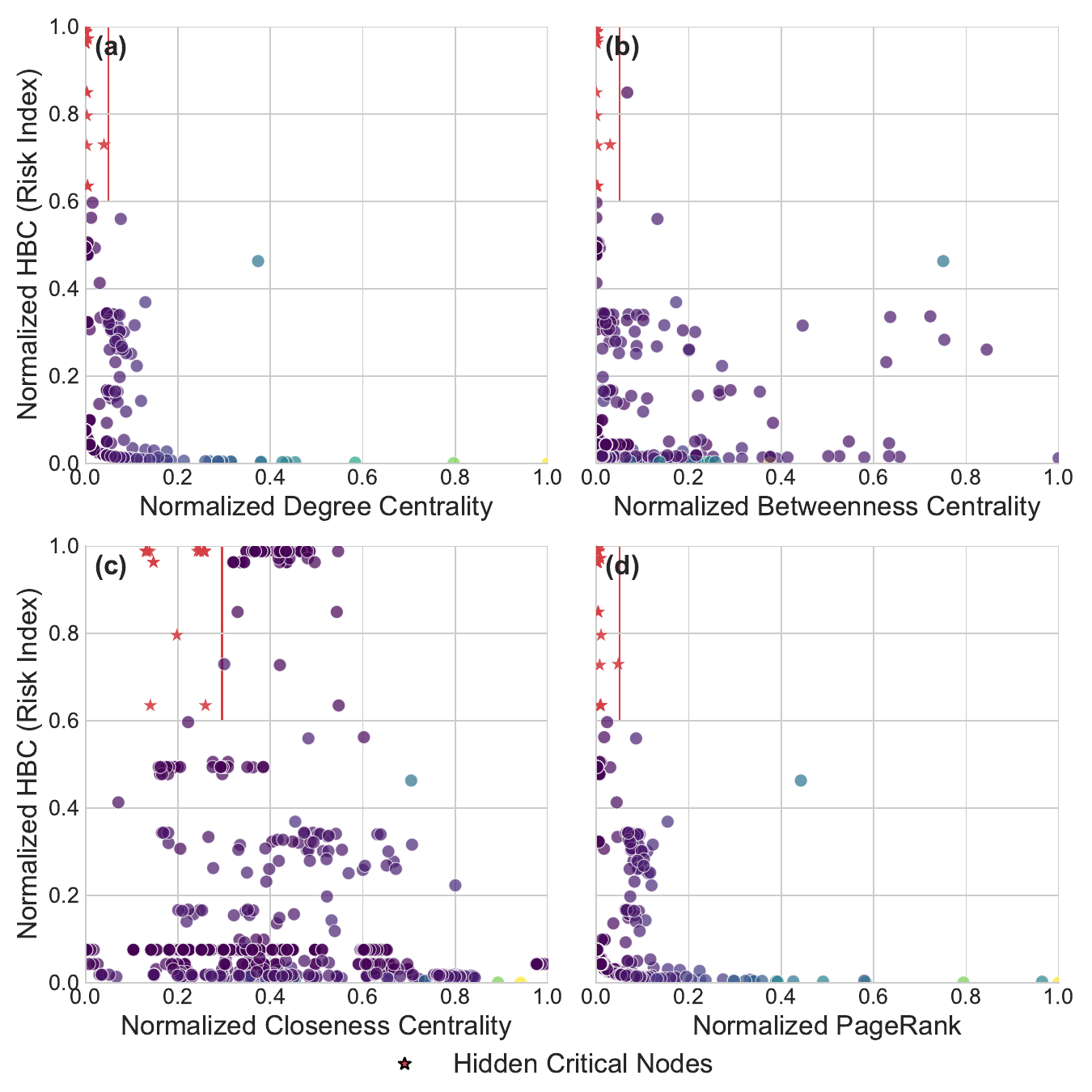}
\caption{\label{rq11} Node distribution by type}
\end{figure}

Figure~\ref{rq11} presents the distribution of nodes across normalized topology-based metrics (x-axis) and HBC (y-axis).
The scatter plot show a distinct cluster of nodes in the top-left region (highlighted in red), these nodes are characterized by low time-averaged centrality (normalized betweenness $< 0.2$) yet exhibit consistently high systemic risk (Avg. HBC $> 0.6$). While they appear insignificant in a static topology map, our hypergraph-based simulation reveals that their failure triggers the most severe network-wide congestion.

\subsection{Characterization of HBC Nodes (RQ2)}
\begin{figure}[t]
\centering
\includegraphics[width=\columnwidth,keepaspectratio]{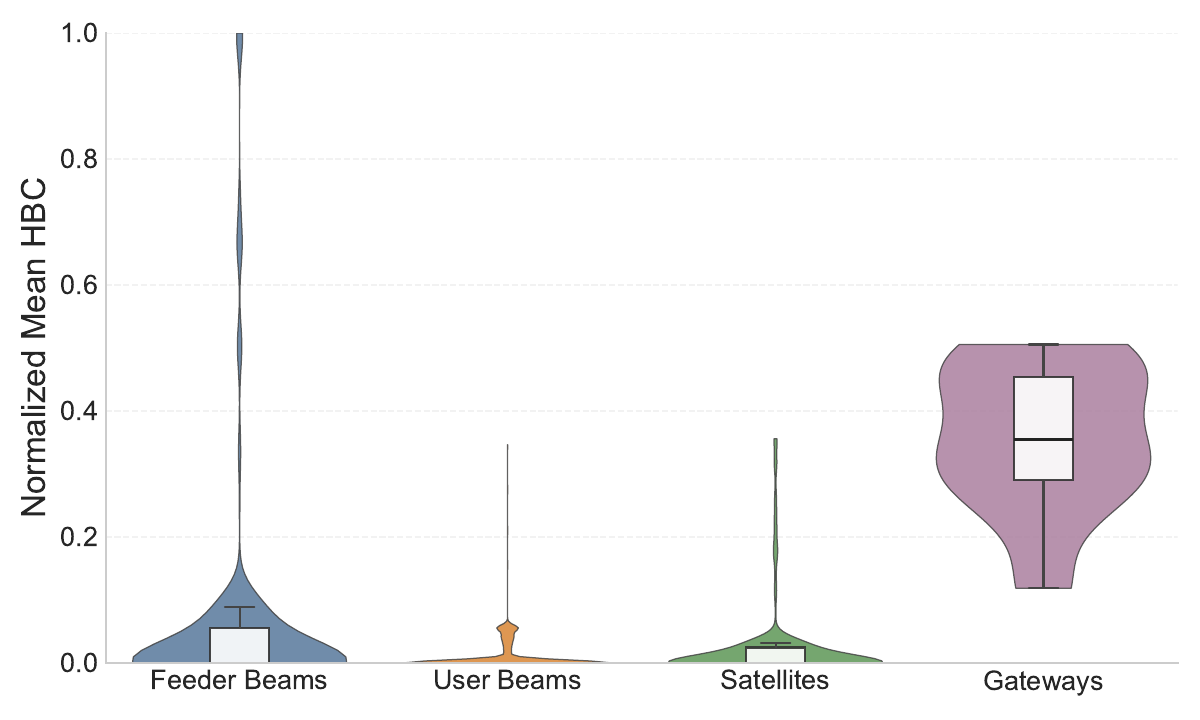}
\caption{\label{rq2} Mean HBC by node type}
\end{figure}
We conducted statistical analysis on the mean HBC values of 4 node types (gateways, satellites, feeder Beams, user Beams) at different time points. 
As illustrated in Figure ~\ref{rq2}, the risk distribution exhibits a pronounced structural hierarchy. Gateways and feeder beams emerge as absolute systemic bottlenecks where network-wide traffic converges. Due to the``one-to-many'' mapping inherent in gateway deployment, these nodes become primary concentration points for systemic risk. Furthermore, feeder beams display significant spatial heterogeneity: while visibility constraints leave most beams idle (zero risk), active ones monopolize massive trunk traffic, resulting in extreme, localized risk peaks.

\begin{figure*}[!t]
\centering
\includegraphics[width=0.9\textwidth,keepaspectratio,trim=5 25 5 30,clip]{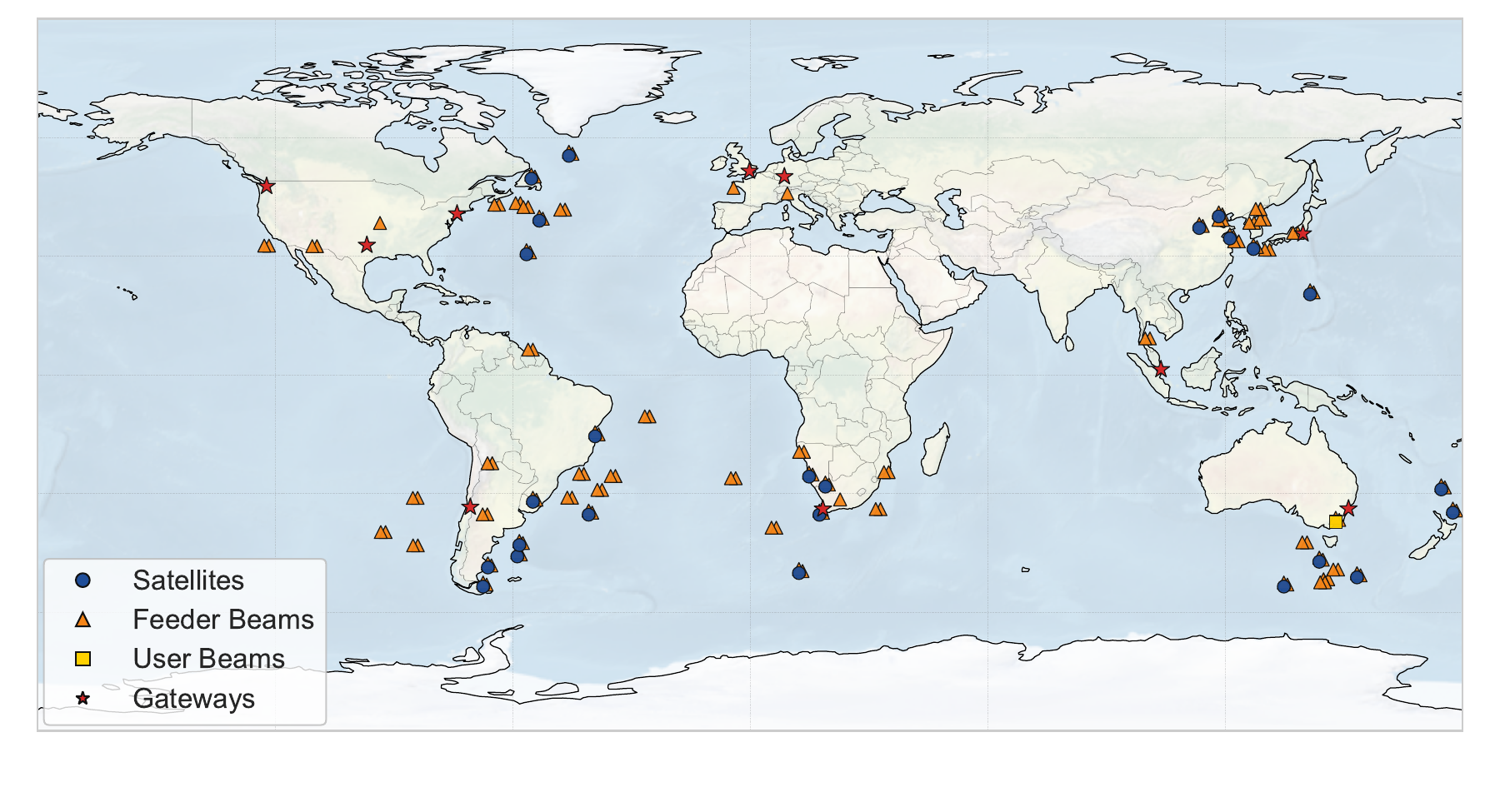}
\caption{\label{rq22} Geographic distribution of top-150 HBC nodes}
\end{figure*}
Figure ~\ref{rq22} characterizes the geographic clustering of high-risk nodes by mapping the top 150 HBC nodes alongside all gateways. The high‑HBC points are not uniformly dispersed; instead, they exhibit clear geographic clustering. This indicates the presence of hidden critical nodes, which temporarily act as non-redundant bridges (e.g., over oceans), disguising high criticality within a traditional statistics. Systemic risk is not solely a function of global constellation uniformity, but is closely tied to the geographic configuration and local density of access/return links.

\subsection{Time-varying Analysis of Risk (RQ3)}

\begin{figure}[t]
\centering
\includegraphics[height=0.25\textheight, keepaspectratio, trim=0 10 0 5, clip]{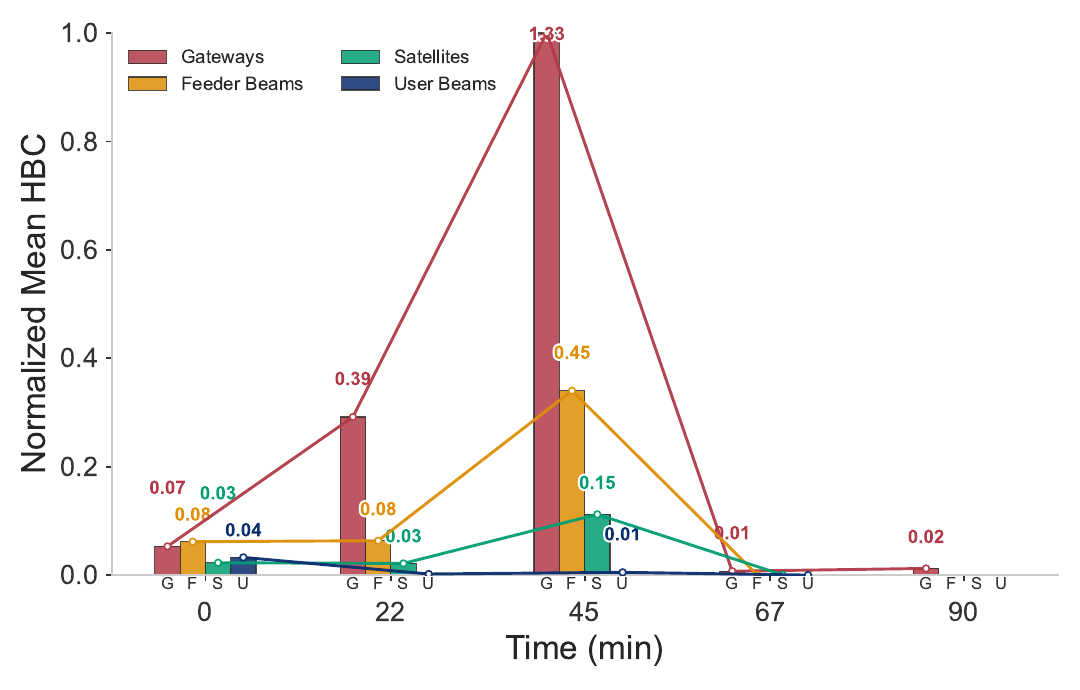}
\caption{\label{rq3} Temporal changes in normalized mean HBC by node type}
\end{figure}

\begin{figure}[t]
\centering
\includegraphics[height=0.25\textheight, keepaspectratio, trim=0 10 0 5, clip]{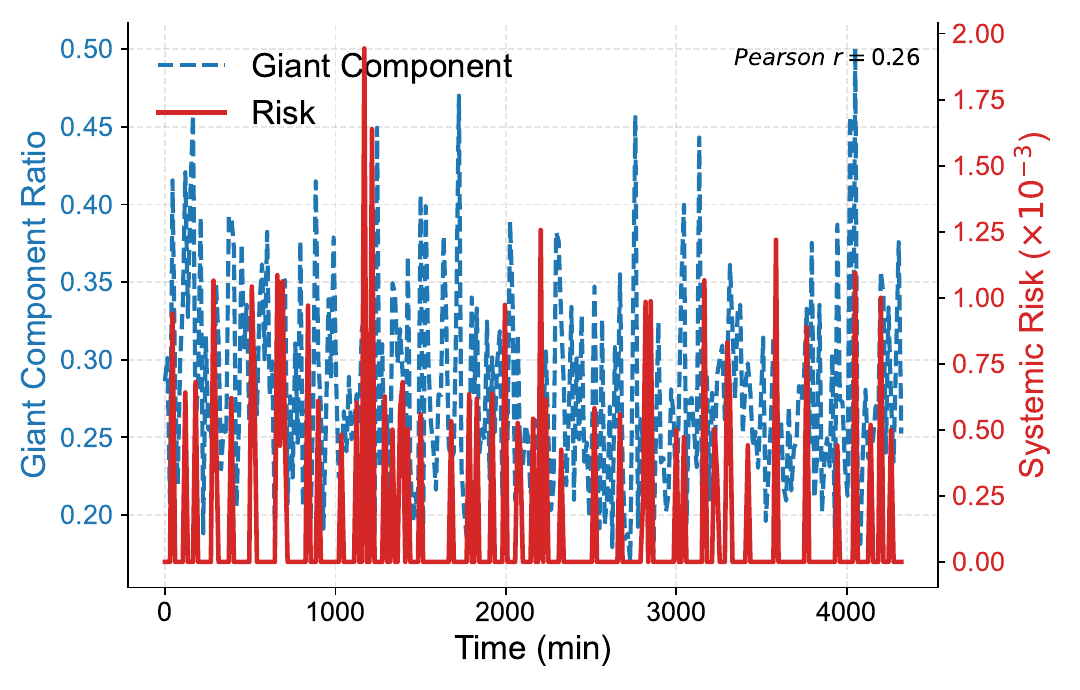}
\caption{\label{rq3_3} Temporal changes in risk and giant component ratio}
\end{figure}
We analyzed the time-varying of HBC by node type, as illustrated in the figure ~\ref{rq3}. We observe a massive singularity where gateway risk surges 19x (from $0.07$ to $1.33$), triggering simultaneous raise in feeder beams ($0.45$) and satellites ($0.15$). This confirms that orbital movement over high-demand regions causes the topology to saturate instantly. The synchronized trajectories of gateways and feeder beams reveal strong back-propagation. Congestion at gateways immediately chokes the upstream feeder links, creating a vertical bottleneck from space to ground. Static metrics mask these critical peaks, a network appearing safe on average can become critically fragile during these predictable orbital orbit variation.

To validate our approach, we extended the simulation to 3 days (4320minutes) with 15-minute sampling. We characterize the temporal dynamics of the constellation using two complementary metrics: the giant component ratio (GCR), which quantifies global structural connectivity, and systemic risk, a metric that captures exposure to link-level congestion. Specifically, systemic risk is defined as the fraction of non-internal links where the projected traffic load, under a fixed-demand regime, exceeds the operational capacity. Our analysis indicates that systemic risk exhibits significant temporal volatility (figure ~\ref{rq3_3}). Notably, the Pearson correlation between GCR and systemic risk is weak ($r = 0.26$), implying that vulnerabilities persist even when the network remains structurally robust.

\subsection{Attack Efficacy and Structural Vulnerability (RQ4)}
We now evaluate the efficacy of HBC against traditional centrality metrics from the perspective of cascading attack impact. 

\textbf{Experimental Setup.} For the attack scenario considered in this study, nodes are sequentially selected as attack targets in increments of 0.02\%, up to a maximum proportion of 2.0\%. We employ varying attack intensity parameters $\alpha$ here, which is given by equation (\ref{eq:tilde_Ci}). ``attack intensity'' does not refer to direct node removal; instead, attacked nodes first undergo capacity degradation, and are only removed when the degradation threshold is met, which is given by equation (\ref{eq:L_i(t)}). To account for the dynamic topology of LSNs, our evaluation spans discrete time slots across a complete orbital cycle, the reported results represent the average value derived from multiple Monte Carlo simulation runs at each time point, We define the Cascading Impact Ratio (CIR) as the ratio of the total number of failed nodes to the number of initially attacked nodes, quantifying the attack’s leverage in triggering cascading  failures.

\begin{figure*}[!t]
\centering
\includegraphics[width=\textwidth, keepaspectratio, trim=5 20 5 30, clip]{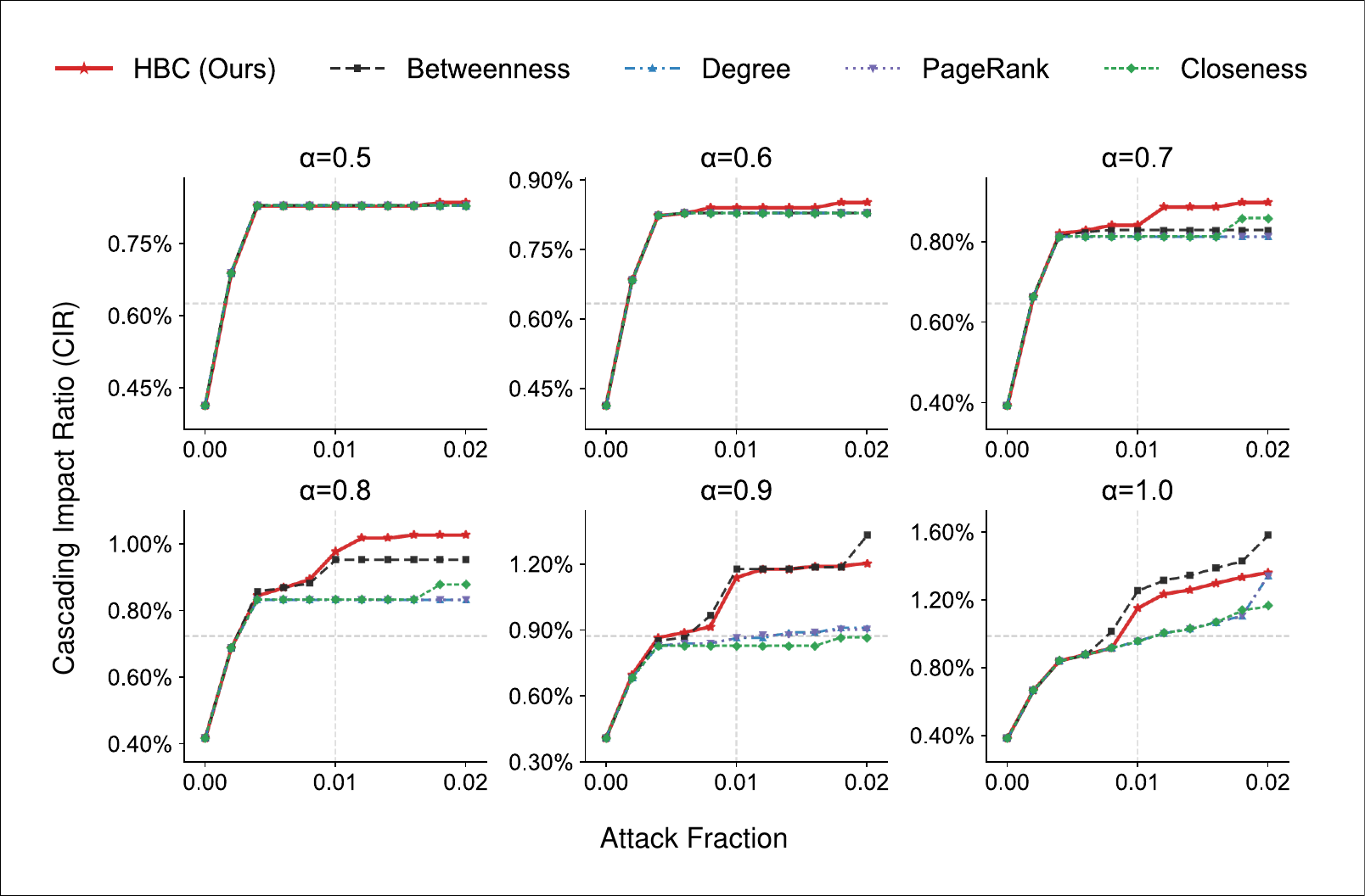}
\caption{\label{rq4} Attack efficacy analysis in Leo constellation}
\end{figure*}

\textbf{Impact of Attack Intensity ($\alpha$).}
The network's response to attacks reveals three distinct phases governed by $\alpha$, as shown in Figure~\ref{rq4}. 

For $\alpha \le 0.6$, the induced load perturbation is insufficient to trigger significant cascades, resulting in negligible performance differences across all metrics. 
Conversely, when $\alpha = 1.0$, the attack intensity is severe enough to cause immediate node failure (bypassing the degradation phase). This effectively reduces the scenario to a direct node removal attack, where the cascading mechanics are less distinct. 

When $\alpha \in [0.7, 0.9]$, is the actual attack strength of the general attack. The network becomes sensitive to specific topological perturbations. In this phase, HBC significantly outperforms all baselines by identifying ``tipping points'' that trigger cascades at lower attack fractions.

\textbf{Efficacy Analysis.} Figure~\ref{rq4} confirms that HBC yields the highest destructive impact. Specifically, in the critical phase ($\alpha=0.7,0.8$), HBC demonstrates a decisive advantage: HBC achieves a substantial 24.1\% relative improvement in cascading impact efficacy compared to local baselines (degree and pagerank). Even against the betweenness centrality, HBC maintains a clear lead, improving the attack efficacy by approximately 8.2\%.

Although an absolute CIR of $\approx 1.0\%$ may appear numerically small, it constitutes a catastrophic disruption in the context of LSNs for two critical reasons: 
First, the inclusion of massive Non-routing user terminals in the CIR denominator dilutes the statistical visibility of core backbone failures. 
Second, our adoption of a conservative active node strategy yields a lower bound of damage compared to worst-case peak-load scenarios (Appendix~\ref{algorithm_detail}). 
Thus, the observed deviation signifies a critical structural vulnerability.

\textbf{Mechanism of ``hidden bridges''.} 
The superior efficacy of HBC can be attributed to its ability to identify ``hidden bridges''—nodes that possess high topological criticality (bridging distinct functional clusters) but exhibit low connectivity (low degree). Unlike high-degree hubs, these hidden bridges are often overlooked by network administrators. However, when these nodes are degraded, traffic is forced to reroute through longer, suboptimal paths, rapidly overwhelming neighbors and propagating congestion more aggressively than attacks on high-centrality hubs.

To further verify the generalizability of our findings, we also conducted extensive experiments to a Walker Delta constellation. Experiment setup and node analysis are provided in Appendix~\ref{sec:appendix_walker}.

\section{Discussion}
In this section, we present several limitations of HYDRA and discuss possible solutions or future works.

\textbf{Computational Complexity.} The theoretical worst-case complexity of the HBC computation is $\mathcal{O}(N^2 \log N)$. By incorporating flow sampling and load-threshold pruning, the search space is effectively constrained, the effective computational complexity in practice is reduced to $\mathcal{O}(N \log N)$. Thus, the proposed algorithm is feasible for on-board computation, enabling distributed, autonomous edge computing without reliance on ground stations. We currently compute risk metrics on ground servers using discrete time snapshots. Our Our offline analysis offers a practical and effective approach to identifying structural weaknesses and informing defense strategy planning.

\textbf{Temporal Interval.}  We initially considered a continuous-time analysis but ultimately decided on discrete snapshots due to the substantial storage overhead it introduced. The basic concept was to capture every micro-second topological change during the orbital period. This would involve monitoring link states after every position update, which resulted in excessive data redundancy. We leave the continuous-time dynamic graph modeling for future work. Our current finding that discrete snapshots effectively capture critical vulnerabilities.

\textbf{Scalability.} The insight of our work (hypergraph modeling of cascade failure) and other strategies are extensible to other LEO constellations like ``OneWeb, Telesat, or Amazon Kuiper''. While our evaluation focused on the Starlink due to its scale and data availability, the physical laws of orbital dynamics apply universally. This framework can be extended to multi-layered constellations, exploring how to relieve the ``Black Swan'' Vulnerabilities identified in this paper.

\textbf{Traffic and Routing Models.} Relying on synthetic traffic models (population-based gravity model) benefits our approach by providing a standardized baseline, but it also constrains the simulation's fidelity. Real-world satellite traffic is highly proprietary and may be influenced by undisclosed algorithms or user special behaviors. Consequently, our method reflects ``potential structural risk'' rather than ``real-time operational load''. Future work requires incorporating real network traces or reverse engineered routing protocols to enhance the accuracy of HYDRA.

\section{Related Work}
\subsection{LSNs Security from Links to Endpoints }

Recent  security assessments in LSNs primarily focused on component level \cite{tedeschi2022}.

At the physical and link layers, studies have addressed ranging security~\cite{coppola}, anti-spoofing~\cite{crosara2024}, and jamming resilience~\cite{yang2025}. Moving to endpoints, researchers have exposed and mitigated vulnerabilities in user terminals~\cite{du2022}, satellite firmware~\cite{willbold2023}, modems~\cite{yu2024 }, and proprietary user protocols~\cite{classen2025}. Additionally, service availability is also threatened by signaling attacks. Liu et al.  introduced SatOver, a cross-layer attack capable of blocking Direct-to-Cell satellites via signaling storms~\cite{liu2024a}. privacy risks regarding user localization via signaling leakage have been rigorously analyzed ~\cite{liu2025, jedermann, koisser2024}. 

However, these studies largely focus on external attacks or single-node hardening. They often overlook the inherent structural fragility of the LSNs routing fabric itself, where local failures can propagate systemically collapse~\cite{yanev2025}.

\subsection{Systemic Resilience and Topology Analysis}
To quantify systemic risks beyond individual nodes, the community adopts complex network theory.

Extensive literature utilizes centrality metrics (Degree, Betweenness) to identify critical nodes~\cite{salim2025}. Hu et al. incorporated temporal graph models to account for dynamic ISL switching~\cite{7510733}. 
However, these approaches typically model LSNs as simple graphs (pairwise edges). This abstraction fails to capture the ``one-to-many'' broadcast nature of satellite beams.

Beyond static topology, the dynamic propagation of failures, where a local perturbation triggers global collapse, is a critical research area. This is often modeled using the Motter-Lai (ML) framework~\cite{motter2002} or epidemic spreading (SIR/SIS) models. In the LEO context, studies have investigated how traffic load redistribution following a failure exceeds the capacity of neighboring satellites, leading to congestion avalanches~\cite{ li2024dynamic}. Researchers have demonstrated bottleneck nodes can severely degrade network throughput~\cite{dengTimevaryingBottleneckLinks2025a}.
Nevertheless, these models rely on simple edge interactions and often neglect the hyper-scale coupling effects introduced by ground-space dynamic change.
\subsection{High-Order and Interdependent Networks}
The limitations of pairwise interaction models have led to the exploration of high-order networks. Theoretical works by Battiston et al. demonstrated that high-order dependencies significantly amplify cascading risks~\cite{2020Networks}. 

In the domain of satellite networks, hypergraphs have been effectively utilized to model complex interference and resource constraints. Hao et al. applied hypergraph structures for cross-domain resource management and spectrum sharing~\cite{hao2022}. Zhang et al. integrated hypergraph neural networks (HGNN) with reinforcement learning to optimize dynamic joint resource allocation~\cite{zhang2025a}. Cao et al.proposed a multi-target robust observation method, maximizing the algebraic connectivity of satellite hypergraphs to maintain tracking precision under failure scenarios~\cite{cao2025}. Similarly, Li et al. modeled LEO satellite hypernetworks to analyze topological attack tolerance~\cite{li2024}.

However, existing studies fail to account for the rapid, time-varying link switching and realistic traffic load distributions inherent to constellations. Bridging this gap, our work introduces a time-varying cascading failure hypergraph framework. By leveraging realistic Starlink TLE data, we unearth hidden structural vulnerabilities that component-level security measures are incapable of mitigating.

\section{Conclusion}
In this paper, we presented HYDRA, a novel framework to unearth systemic vulnerabilities hidden within the dynamic topology of LSNs. By introducing a cascade failure hypergraph abstraction, our approach mathematically captures the ground-space bottleneck nature and orbital dynamics inherent to satellite communications.

First, our analysis reveals a significant divergence between topological centrality and structural criticality; we identify that the nodes essential for maintaining global connectivity are often not the highly connected central hubs, but rather the topologically sparse ``Oceanic Bridges'' located at mid-latitudes. Lacking redundant paths, these nodes act as ``Black Swan'' nodes that lead to network colleaspe, yet they are frequently underestimated by traditional topolog-based assessments. 

Second, systemic risk exhibits pronounced temporal volatility. Driven by the coupling of orbital dynamics and ground traffic distributions, network risk is not a static constant but manifests as time-vary and periodic, demonstrating that risk assessment methods relying solely on static snapshots hard to capture transient vulnerabilities during operation.  

Finally, the proposed HBC metric demonstrates superior efficacy in identifying critical vulnerabilities. Quantitatively, small-scale targeted attacks based on HBC result in the average highest cascading impact ratio. Specifically, HBC outperforms degree-based strategies by a substantial relative margin of 24.1\% and exceeds the robust Betweenness Centrality by approximately 8.2\%. While these numerical deviations are inherently diluted by the massive scale of user terminals, they signify a disproportionately high failure rate among core routing nodes. This precise identification of hidden weaknesses provides a more effective theoretical basis for designing redundancy strategies against cascading failures.

In conclusion, our findings collectively underscore the necessity of integrating  physical-layer orbital characteristics with network topology in LSNs security analysis, catalyzing a paradigm shift from static topological analysis to dynamic structural analysis that aligns with the physical reality of LSNs.

\appendix
\section*{Acknowledgments}
This work was supported by the National Key Research and Development Program of China (Grant No. 2022YFB4501000), the National Natural Science Foundation of China (Grant Nos. 62271019, 62225201, and 62202021), and the National Cyber Security-National Science and Technology Major Project.

\section*{Open Science}
To support open science and facilitate reproducibility, we have implemented the proposed HYDRA framework and made the source code publicly available. Additionally, the repository includes the full simulation datasets for Starlink TLE used in our evaluation. The anonymized artifacts for the review process can be accessed at:\url{https://github.com/Tobin-BUAA/LEO-Satellite-HBC-Sim}.

\bibliographystyle{plain}
\bibliography{jobname}

\section{Simulation Framework Overview}
The simulator constructs a time‑varying ground-space network to study cascading failures, criticality, and defense strategies. Inputs include Starlink TLEs, gateway/population data, and configuration parameters (e.g., ISL range, elevation threshold, target load). Outputs consist of connectivity, risk, and performance indicators, saved as CSV for reproducibility and analysis.

\textbf{Orbital Propagation \& Coordinates.} Satellite positions are propagated by SGP4 at discrete time steps. TEME coordinates are rotated to ECEF using GMST, nabling consistent geometry with ground nodes. Positions are wsed for visibility checks, link distances, and delays.

\textbf{Multi‑layer Topology Construction.} Nodes include satellites, gateways, user-beams, feeder-beams, and users. Edges include ISL, feeder, access, and internal beam-sat links. Each node/edge carries capacity and delay attributes for subsequent load and failure modeling.

\textbf{Traffic Generation \& Load Scaling.} User flows are sampled by adoption weights. Demand scales with population weight and a local time factor. Demands are globally normalized to match target load and effective capacity.

\textbf{Load‑Aware Routing \& Cascades.} Routing uses load-aware Dijkstra with weights inversely proportional to residual capacity plus delay. Overload of nodes or links triggers failure; cascading proceeds iteratively until stable or max iter is reached.

\section{Active Node Selection}
\label{algorithm_detail}
Active satellites are chosen by (i) user visibility, (ii) gateway visibility, (iii) feeder feasibility, and (iv) optional neighborhood expansion.
For each user, visible satellites above the elevation threshold are ranked by elevation (desc) and distance (asc), and top‑(K) are selected.
Gateways contribute their top‑($K_g$) nearest visible satellites. Satellites that can form feeder links are retained.
If the “expand” mode is enabled, the active set is expanded via k‑NN neighbors within a maximum ISL distance over multiple hops.
\begin{algorithm}[!htbp]
\caption{Active Satellite Selection}
\label{alg:active_sat_selection}
\KwIn{All satellites $\mathcal{S}$, users $\mathcal{U}$, gateways $\mathcal{G}$}
\KwOut{Active satellite set $\mathcal{S}_a$}

$\mathcal{S}_a \leftarrow \emptyset$\;
\ForEach{$u \in \mathcal{U}$}{
  Find candidate satellites with elevation $\ge \theta_{\min}$\;
  Rank candidate satellites by (elevation descending, distance ascending), and take top-$K$\;
  $\mathcal{S}_a \leftarrow \mathcal{S}_a \cup \text{selected satellites}$\;
}

$\mathcal{S}_g \leftarrow$ top-$K_g$ visible satellites for each gateway\;
$\mathcal{S}_a \leftarrow \mathcal{S}_a \cup \mathcal{S}_g$\;

\ForEach{$s \in \mathcal{S}_a$}{
  \If{any gateway is visible to satellite $s$}{
    Assign feeder beams to satellite $s$ and keep $s$ in $\mathcal{S}_a$\;
  }
}

\If{EXPAND mode is enabled}{
  Expand $\mathcal{S}_a$ by k-NN neighbors within $D_{\max}$ for $H$ hops\;
}

\Return{$\mathcal{S}_a$}
\end{algorithm}

\section{Evaluation on Idealized Walker Constellations}
\label{sec:appendix_walker}
We constructed an idealized Walker Delta constellation for comparison experiments.
The specific configuration parameters are detailed in Table ~\ref{tab:walker_params}. 
In this setup, satellite coordinates are generated analytically using a two-body approximation, while the traffic generation, link establishment rules, and cascading dynamics remain identical to the primary experiment. 
\begin{table}[htbp]
    \centering
    \caption{Configuration parameters for the idealized Walker Delta constellation}
    \label{tab:walker_params}
    \begin{tabular}{p{0.35\columnwidth} p{0.55\columnwidth}}
        \toprule
        \textbf{Parameter} & \textbf{Value / Description} \\
        \midrule
        Inclination ($i$) & $53^\circ$ \\
        Total Satellites ($T$) & $1584$ \\
        Orbital Planes ($P$) & $72$ \\
        Satellites per Plane ($T/P$) & $22$ \\
        Phasing Factor ($F$) & $39$ \\
        Altitude ($h$) & $550$ km \\
        Link Constraints & 4 ISLs (2 intra-plane, 2 inter-plane) \\
        \bottomrule
    \end{tabular}
\end{table}

We also investigate whether the proposed HBC metric can identify critical nodes that are overlooked by classical topology-based metrics.
\begin{figure}[t]
\centering
\includegraphics[width=\columnwidth,keepaspectratio,trim=2 2 2 2,clip]{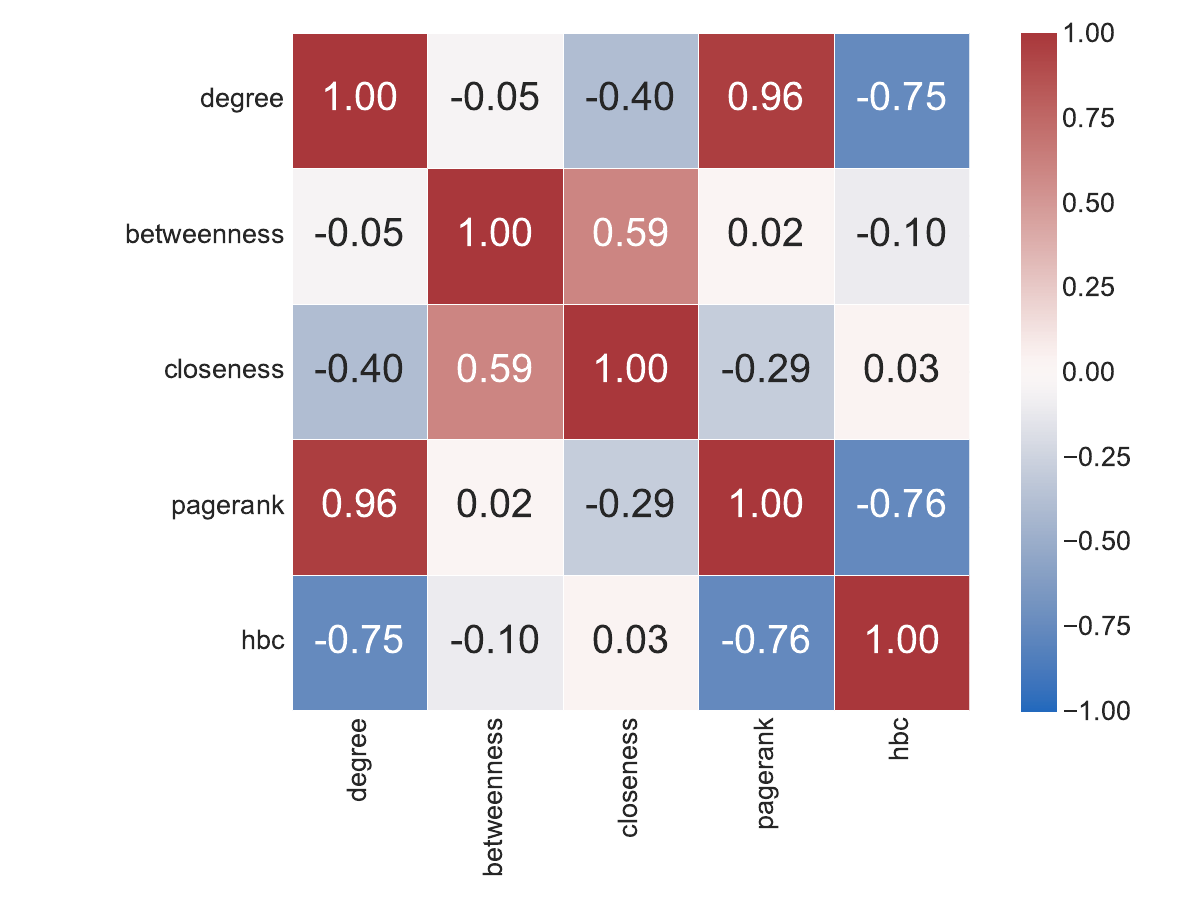}
\caption{\label{cw} Pearson correlation matrix of metrics (Walker Delta constellation)}
\end{figure}
Figure~\ref{cw} presents the Walker Delta constellation Pearson correlation matrix among metrics. This indicates a strong negative correlation between HBC and other metrics, confirming their lack of correspondence.

\begin{figure}[!t]
\centering
\includegraphics[width=\columnwidth,keepaspectratio,trim=2 2 2 2,clip]{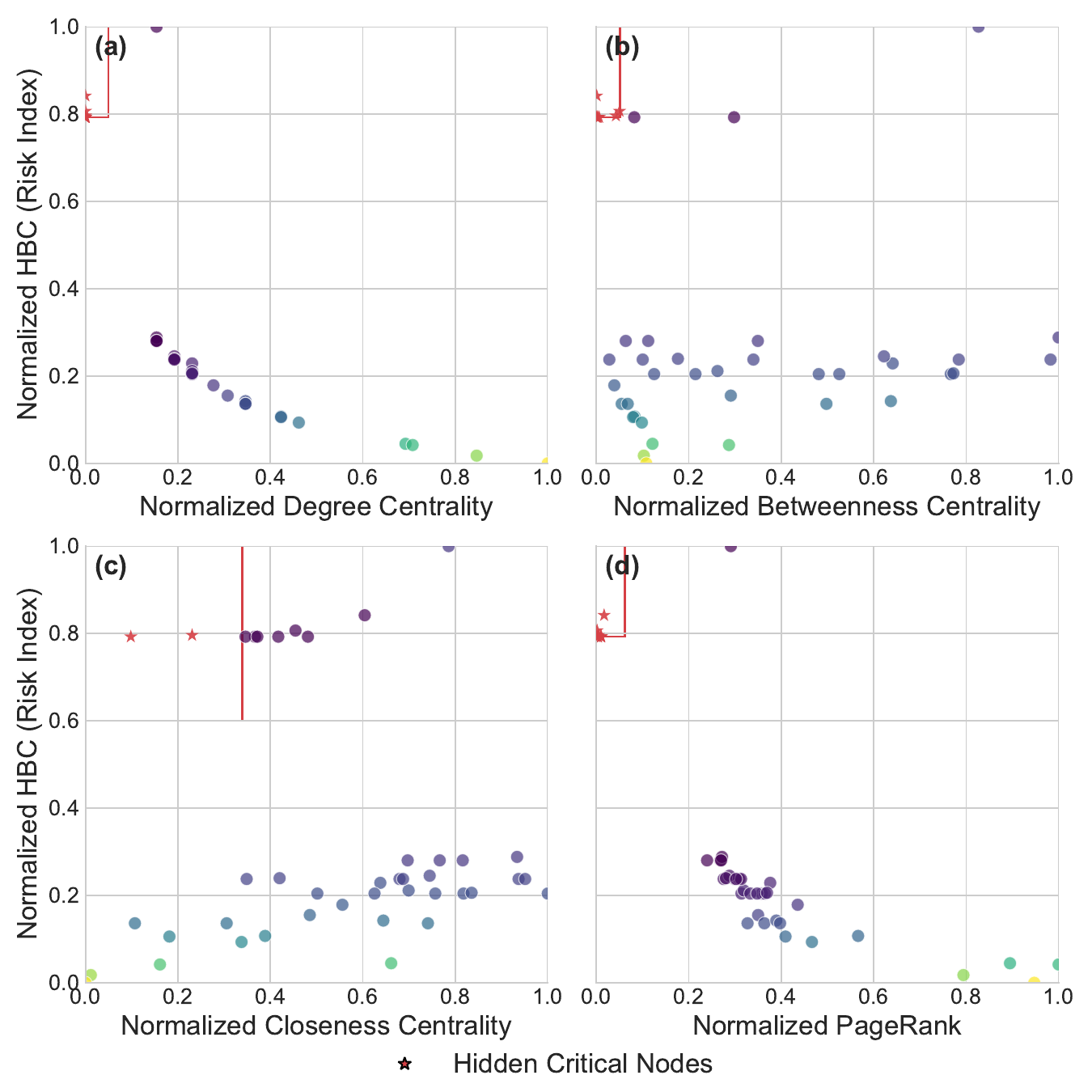}
\caption{\label{hw} Node distribution by type (Walker Delta constellation)}
\end{figure}

Figure~\ref{hw} illustrates the divergent distribution of HBC relative to other topological metrics and the consequent difficulty in identifying such structurally critical nodes via conventional topology-based indicators.
\begin{figure*}[!htbp]
\centering
\includegraphics[width=\textwidth, keepaspectratio, trim=5 20 5 30, clip]{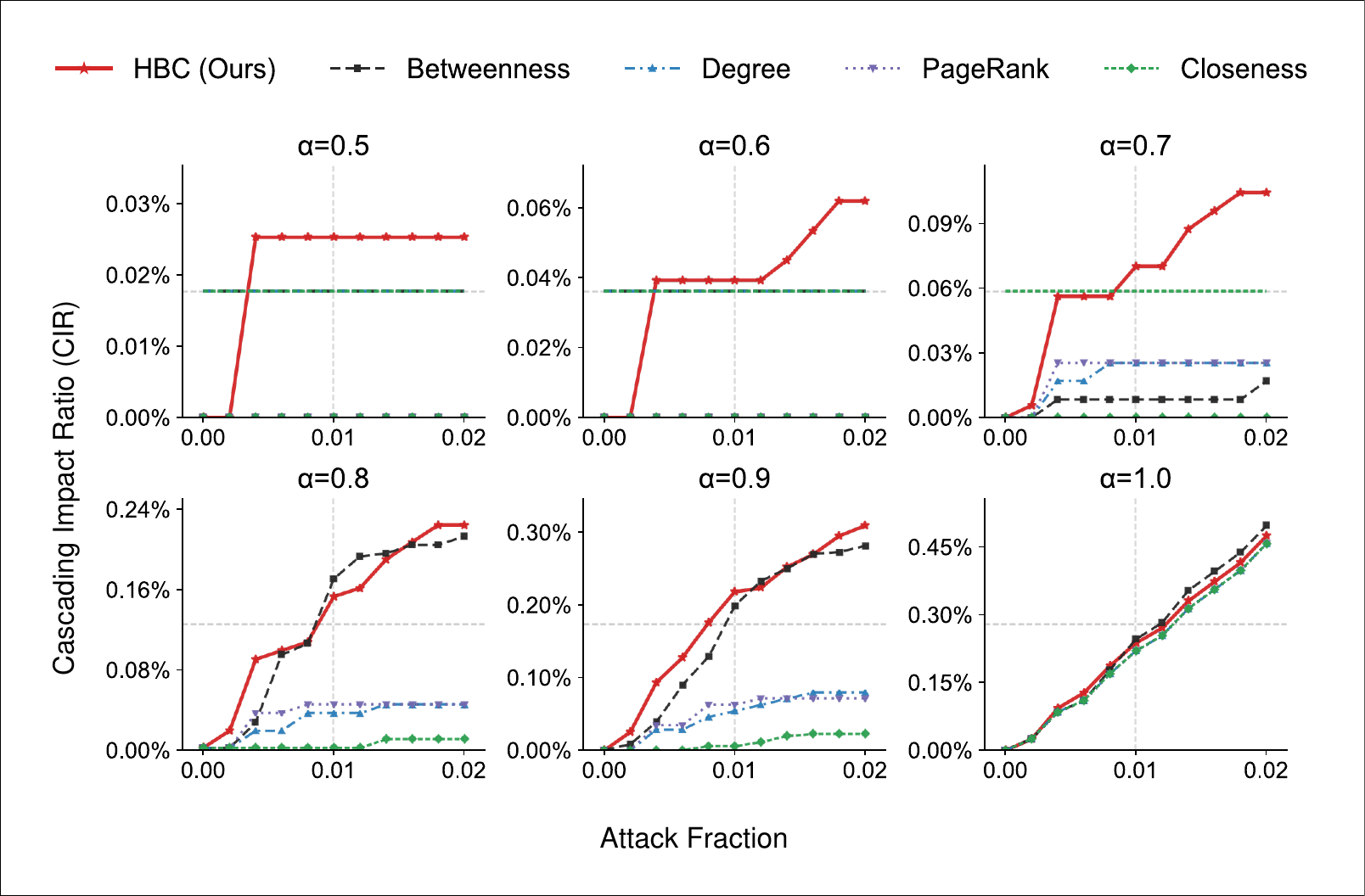}
\caption{\label{rq44} Attack efficacy analysis in Walker Delta constellation}
\end{figure*}

As shown in Figure~\ref{rq44}, The results strongly corroborate the superior sensitivity of HBC, particularly in the early stages of cascading failure. In the critical initiation phase ($\alpha=0.7$), standard centrality metrics (e.g., Betweenness and Degree) fail to trigger significant disruptions, in stark contrast, HBC successfully identifies hidden structural vulnerabilities. This represents a $>$5-fold relative improvement in attack efficacy, demonstrating that HBC is significantly more agile in exposing the fragility of diverse satellite network architectures before other metrics even register a threat.

\end{document}